\documentclass[12pt,eadjoint tfnpsf]{article}

\catcode`\@=11
\@addtoreset{equation}{section}

\global\arraycolsep=1pt

\usepackage{amsbsy,amssymb,latexsym,amsfonts}
\usepackage{mathrsfs}
\usepackage{graphicx}
\RequirePackage[dvips,usenames]{color}
\definecolor{fireblick}{rgb}{0.698039,0.133333,0.133333}

\newcommand{\beq}{\begin{equation}}
\newcommand{\eeq}{\end{equation}}
\newcommand{\bea}{\begin{eqnarray}}
\newcommand{\eea}{\end{eqnarray}}

\newcommand{\CF}{{\mathcal F}}

\newcommand{\CN}{{\mathcal N}}

\newcommand{\CW}{{\mathcal W}}

\renewcommand\Im{{\mathrm{Im}}}

\def\Tr{\mathop{\rm Tr}}
\newcommand\tr{\mathrm{tr}}

\newcommand\diag{\mathrm{diag}}

\begin{document}
%%%%%%%%%%%%%%%%%%%%%%%%%%%%%%%%%%%%%%%%%%%%%%%%%%%%%%%%%%%%%%%%%%%%%%%%%%%%%%%%%%%%%%%%%%
%
% title page
%
%%%%%%%%%%%%%%%%%%%%%%%%%%%%%%%%%%%%%%%%%%%%%%%%%%%%%%%%%%%%%%%%%%%%%%%%%%%%%%%%%%%%%%%%%%
\begin{titlepage}
%%%%%%%%%%%%%%%%%%%% preprint # %%%%%%%%%%%%%%%%%%
\begin{flushright}
\normalsize
%\filename
~~~~
YITP-09-26\\
April, 2009 \\
\end{flushright}
%%%%%%%%%%%%%%%%%%%%%%%%%%%%%%%%%%%%%%%%%%%%%%%%%%

\vspace{72pt}

%%%%%%%%%%%%%%%%%%%% title %%%%%%%%%%%%%%%%%%%%%%%
\begin{center}
{\Large Quiver Gauge Theory and Extended Electric-magnetic Duality}\\
\end{center}
%%%%%%%%%%%%%%%%%%%%%%%%%%%%%%%%%%%%%%%%%%%%%%%%%%

\vspace{25pt}

%%%%%%%%%%%%%%%%%%% authors %%%%%%%%%%%%%%%%%%%%%%
\begin{center}
{%
Kazunobu Maruyoshi\footnote{e-mail: maruyosh@yukawa.kyoto-u.ac.jp}
}\\
%%%%%%%%%%%%%%%%%%%%%%%%%%%%%%%%%%%%%%%%%%%%%%%%%%
%
\vspace{15pt}
%
%%%%%%%%%%%%%%%%%%% affiliation %%%%%%%%%%%%%%%%%%%
\it Yukawa Institute for Theoretical Physics, Kyoto University, Kyoto 606-8502, Japan\\
\end{center}
%%%%%%%%%%%%%%%%%%%%%%%%%%%%%%%%%%%%%%%%%%%%%%%%%%%
%
\vspace{20pt}
\begin{center}
Abstract\\
\end{center}
%%%%%%%%%%%%%%%%%%%% abstract %%%%%%%%%%%%%%%%%%%%%
  We construct $\CN=1$ $A$-$D$-$E$ quiver gauge theory with the gauge kinetic term 
  which depends on the adjoint chiral superfields, 
  as a low energy effective theory on D5-branes wrapped on 2-cycles of Calabi-Yau 3-fold in IIB string theory.
  The field-dependent gauge kinetic term can be engineered by introducing $B$-field 
  which holomorphically varies on the base space (complex plane) of Calabi-Yau.
  We consider Weyl reflection on $A$-$D$-$E$ node, which acts non-trivially on the gauge kinetic term.
  It is known that Weyl reflection is related to $\CN=1$ electric-magnetic duality.
  Therefore, the non-trivial action implies an extension of the electric-magnetic duality 
  to the case with the field-dependent gauge kinetic term.
  We show that this extended duality is consistent from the field theoretical point of view.
  We also consider the duality map of the operators.
%%%%%%%%%%%%%%%%%%%%%%%%%%%%%%%%%%%%%%%%%%%%%%%%%%%

\vfill

\setcounter{footnote}{0}
\renewcommand{\thefootnote}{\arabic{footnote}}

\end{titlepage}

%%%%%%%%%%%%%%%%%%%%%%%%%%%%%%%%%%%%%%%%%%%%%%%%%%%%%%%%%%%%%%%%%%%%%%%%%%%%%%%%%%%%%%%%%%%%%%%%%%% section 1
\section{Introduction}
\label{sec:intro}
  In the previous decade, various exciting investigations have been made on 4d, $\CN=1$ supersymmetric gauge theory.
  In string theory point of view, supersymmetric gauge theory can be realized 
  as a low energy effective theory on D-branes.
  The low energy behavior of supersymmetric gauge theory has been widely analyzed by using gauge/gravity correspondence.
  In particular, it has been known that the effective superpotential of $\CN=1$ supersymmetric gauge theory 
  with an adjoint chiral superfield and a tree level superpotential can be evaluated
  from the gravity theory with flux \cite{Vafa, CIV, CFIKV} and from the matrix model \cite{DV}.
  These relations have been analyzed in \cite{DGLVZ, CDSW, Seiberg2, CSW2} field-theoretically.
  
  Recently, some interesting results have been obtained in $\CN=1$ supersymmetric $U(N)$ gauge theory
  with the gauge kinetic term which depends on the adjoint chiral superfield, 
  $\Im \int d^2 \theta \Tr \tau(\phi) \CW^\alpha \CW_\alpha$.
  In \cite{IM, Ferrari}, it has been shown that the effective superpotential of such a theory
  is deformed compared to the theory with constant $\tau$
  (in \cite{IM}, a specific case where $\CN=2$ supersymmetry is spontaneously broken to $\CN=1$ \cite{FIS, IMS}
  has been analyzed).
  Since there are non-renormalizable coupling constants, this theory should have UV completion.
  In \cite{ABSV}, this theory is obtained as a low energy effective theory on D5-branes
  wrapped on $S^2$'s in Calabi-Yau 3-fold.
  The field-dependent gauge kinetic term is introduced by the integrals over $S^2$'s 
  of non-trivial $B$-field flux which holomorphically varies on the base space (complex plane) of Calabi-Yau.
  By using geometric transition duality, 
  the deformed superpotential \cite{IM, Ferrari} has been derived from the gravity theory \cite{ABSV}.
  Also, it has been argued that the deformation of the effective superpotential
  leads to the existence of supersymmetry breaking vacua in some cases of the parameters.
  (See also \cite{OT} for IIA and M-theory perspectives 
  and \cite{Maruyoshi} for the case with partially broken $\CN=2$ supersymmetry.)
  
  In this paper, we study $\CN=1$ supersymmetric gauge theory with the flavors
  where the gauge kinetic term depends on the adjoint chiral superfields.
  In the first half of the paper, we analyze $\CN=1$ $A$-$D$-$E$ quiver gauge theories.
  As in \cite{CKV, CFIKV, OTatar}, 
  $\CN=1$ $A$-$D$-$E$ quiver gauge theories can be obtained as low energy effective theories
  on D5-branes wrapped on $S^2$'s in Calabi-Yau 3-folds (and D3-branes in affine case) in IIB string theory.
  More precisely, these Calabi-Yau 3-folds are constructed by fibering the non-singular spaces, 
  which are obtained from the ALE spaces with $A$-$D$-$E$ singularity, over the complex plane $t$.
  As in \cite{ABSV} for $A_1$ case, the non-trivial $B$-field flux which depends on $t$ is turned on 
  in order to obtain the field-dependent gauge kinetic term.
  
  An interesting result of the string theory construction of $A$-$D$-$E$ quiver gauge theories is that
  the theory which is obtained by Weyl reflection on a node of the Dynkin diagram (or equivalently quiver diagram) 
  is equivalent to the original one,
  due to ambiguities from the fact that there is no unique way to blow up the singularity.
  Since the nodes of the Dynkin diagrams correspond to $S^2$'s, Weyl reflection acts on the gauge kinetic term 
  as well as the superpotential in the gauge theory. 
  In \cite{KlebanovS, CFIKV}, it has been analyzed, in the constant $\tau$ case, that
  the theory obtained by Weyl reflection is a dual description
  (by Kutasov duality \cite{Kutasov, ASY, KS, KSS, EGKRS}),
  after integrating out the meson fields and the flavors.
  (See also \cite{ABSV0, ABF} for the case with antibranes.)
  
  In the case which we will consider in this paper, the gauge kinetic term is affected by Weyl reflection.
  Therefore, we propose an extension of the Kutasov duality:
  $\CN=1$ supersymmetric $U(N_c)$ gauge theory with an adjoint $\phi$ and $N_f$ flavors $Q$ and $\bar{Q}$,
  equipped with the gauge kinetic term where $\tau(\phi)$ is
    \bea
    \tau(\phi)
     =     \sum_{k = 0}^m t_k \phi^k,
    \eea
  and a superpotential, 
  has a dual description which is $U(n N_f - N_c)$ gauge theory 
  with an adjoint $\tilde{\phi}$, $N_f$ flavors $q$ and $\bar{q}$ and meson fields,
  where the gauge kinetic term is 
    \bea
    \tilde{\tau}(\tilde{\phi})
     =   \tilde{t}_0 - \sum_{k = 1}^m t_k \tilde{\phi}^k
    \eea
  and the dual superpotential is the same as the one in \cite{Kutasov, KS, KSS}.
  In the latter half of this paper, we will analyze this duality
  from the field theoretical point of view.
  
  The dual superpotential and the dual gauge kinetic term can be determined by the consistency of the duality.
  In particular, a strong constraint is 
  that both theories should split to decoupled SQCD theories at low energy 
  and there exists a corresponding SQCD theory in the dual theory which is Seiberg dual \cite{Seiberg} 
  to each decoupled SQCD theory in the original one.
  The duality map of several operators can also verified 
  by using the above argument and the generalized Konishi anomaly equations.
  (See also \cite{Ferrarif} for a different analysis of this theory.)
  
  The organization of this paper is as follows.
  We introduce $\CN=1$ $A$-$D$-$E$ quiver gauge theories with the gauge kinetic terms 
  which depend on the adjoint chiral superfields, promoting the gauge coupling constants $\tau_i$ 
  to the field-dependent functions $\tau_i (\phi_i)$, in section \ref{sec:classical}.
  We will show that the classical equations of motion reduce to the same vacuum equations as those in the theories 
  with constant $\tau_i$.
  In section \ref{sec:geometric}, we construct such theories in the framework of superstring theory.
  We then consider a duality: Weyl reflection on $A$-$D$-$E$ nodes in section 4.
  We will see that this induces a non-trivial action on the gauge kinetic term as well as the superpotential.
  In section \ref{sec:fieldtheory}, we consider a non-trivial check of the duality proposal.
  Finally, we will analyze the duality map of the operators, in section \ref{sec:dualitymap}.

%%%%%%%%%%%%%%%%%%%%%%%%%%%%%%%%%%%%%%%%%%%%%%%%%%%%%%%%%%%%%%%%%%%%%%%%%%%%%%%%%%%%%%%%%%%%%%%%%%% section 2
\section{$A$-$D$-$E$ quiver gauge theories}
\label{sec:classical}
  In this section, we introduce $\CN=1$ $A$-$D$-$E$ quiver gauge theories.
  Throughout this paper, we consider the case where the gauge kinetic term depends on the adjoint chiral superfields. 
  This is an extended version of the quiver gauge theories considered in \cite{CKV, CFIKV, OTatar}.
  
  These theories are considered as a deformation of $\CN=2$ $A$-$D$-$E$ quiver gauge theories
  whose gauge groups are $\prod_i U(N_i)$ and each gauge factor corresponds to each node of the quiver diagrams.
  The quiver diagrams are expressed by the Dynkin diagrams of non-affine or affine $G = A, D, E$ groups.
  In terms of $\CN=1$ superfields, $\CN=2$ quiver gauge theory consists of the vector superfields $V_i$ 
  (or the field strength superfields $\CW^\alpha_i$),
  the adjoint chiral superfields $\phi_i$ and the matter chiral superfields $Q_{ij}$ and $Q_{ji}$ which
  are respectively in the bi-fundamental representations $(N_i, \bar{N}_j)$ and $(\bar{N}_i, N_j)$ 
  of $U(N_i) \times U(N_j)$ groups.
  ($i$ label the nodes of the quiver diagram.)
  We consider the case where the prepotential 
  which determines the $\CN=2$ classical Lagrangian has higher order terms, that is, 
    \bea
      \CF_i (\Psi_i)
       =     \sum_{k=0} \frac{t_{i, k}}{(k +1)(k +2)} \Psi_i^{k + 2},
    \eea
  where $\Psi_i$ are the $\CN=2$ vector superfields which contain $\phi_i$ and $\CW_i^\alpha$ 
  and $t_{i, k}$ are complex parameters.
  In $\CN=1$ superspace formalism, this leads to the field-dependent gauge kinetic term
    \bea
    \sum_i \Im \int d^2 \theta \Tr \tau_i (\phi_i) \CW^\alpha_i \CW_{i \alpha},
    \eea
  and also the K\"ahler terms.
  Here $\tau_i$ are related with the prepotentials as $2 \tau_i (x) = \CF''_i (x)$.
  
  We add the superpotentials $W_i (\phi_i)$ which break $\CN=2$ supersymmetry to $\CN=1$.
  We will choose these superpotentials to be polynomials of the same degree $n + 1$, for simplicity.
  Therefore, the holomorphic part of the Lagrangian is
    \bea
    \sum_i \left[ \Im \int d^2 \theta \Tr \tau_i (\phi_i) \CW^\alpha_i \CW_{i \alpha}
    + \int d^2 \theta \left( \Tr \sum_j s_{ij} Q_{ij} Q_{ji} \phi_i - \Tr W_i (\phi_i) \right)
    + h.c. \right],
    \eea
  where $s_{ij}$ is the intersection matrix of $i$-th and $j$-th nodes, which is zero if the nodes are not linked 
  and $\pm 1$ if linked (and they also satisfy $s_{ij} = - s_{ji}$).
  In the affine case, the following condition for the superpotentials:
    \bea
    \sum_{i = 0}^r d_i W_i (x)
     =     0
    \eea
  should be satisfied (where $d_i$ are the Dynkin indices), 
  if we geometrically engineer this theory \cite{CKV}.
  In the non-affine case, there is no restriction on the superpotentials.
  Note that in the case where $t_{i, k}= 0$ for $k>0$, i.e. constant $\tau_i$ case, 
  these theories reduce to the ones analyzed in \cite{CKV, CFIKV, OTatar}.
  
%%%%%%%%%%%%%%%%%%%%%%%%%%%%%%%%%%%%%%%%%%%%%%%%%%%%%%%%%%%  subsection 2.1
\subsection{Classical equations of motion}
\label{subsec:eom}
  The solution of the classical equations of motion 
  in the case where $t_{i, k}= 0$, for $k>0$ (constant $\tau_i$ case) has been derived in \cite{CKV}.
  In non-zero $t_{i, k}$ case, the equations of motion are slightly complicated, 
  but reduce to the same conditions as those in \cite{CKV}.
  Let us see this in this subsection.
  
  First of all, it is easy to see that the F-term equations are not changed 
  compared to the case with constant $\tau_i$.
  In fact, the gauge kinetic term which depends on the adjoint chiral superfields induces an additional term,
  $\partial_{\phi_i} \tau_i (\phi_i)$ multiplied by the fermion bilinear, 
  in the F-term equations with respect to $\phi_i$.
  However, the fermion does not get an expectation value in the classical vacua which we are interested in. 
  Therefore, this term does not contribute to the solution.
  
  On the other hand, the equations of motion with respect to $D^a_i$ 
  (where $a = 1, \ldots, N^2_i$ label the gauge indices of $U(N_i)$ gauge group) lead to
    \bea
    D^a_i
     =     \frac{i}{2} (f_i)_{~bc}^a \bar{\phi}_i^b \phi_i^c
         - (\Im \tau_i (\phi_i)^{-1})^{ab} 
           \left[ \Tr \sum_{j} s_{ij} (Q_{ij}^\dag t_b^i Q_{ij} - Q_{ji} t_b^i Q^\dag_{ji}) + h.c. \right]
     =     0,
           \label{D}
    \eea
  where $(f_i)^{a}_{~bc}$ and $t_a^i$ are the structure constants and the generators in the fundamental of $U(N_i)$.
  Each matrix $(\Im \tau_i (\phi_i)^{-1})^{ab}$ is defined as an inverse of $N^2_i \times N_i^2$ matrix
  $\Tr (\Im \tau_i (\phi_i) t_a^i t_b^i)$.
  While we have non-trivial factors $(\Im \tau_i (\phi_i)^{-1})^{ab}$ in (\ref{D})
  compared to the case with constant $\tau_i$ (in this case, the factors are proportional to $\delta^{ab}$),
  (\ref{D}) results in
    \bea
    \sum_j s_{ij} (Q_{ij} Q_{ij}^\dag - Q^\dag_{ji} Q_{ji} ) + h.c.
     =     0,
           \label{D2}
    \eea
  which are the same D-term conditions as those in the constant $\tau_i$ case.
  This can be seen as follows.
  We are interested in the vacua where the scalar fields get the diagonal vev, 
  i.e. non-Cartan parts of $\langle \phi_i \rangle$ are zero.
  Under these, the first term of (\ref{D}) is zero.
  Since $\det ( \Im \tau_i (\phi_i)^{-1} )^{ab} \neq 0$, the solution is trivial and we obtain (\ref{D2}).
  
  We have shown that the vacuum conditions following from the equations of motion 
  are the same as those in \cite{CKV}.
  Therefore, the structure of the classical vacua is also same.
  So, we only explain these here.
  
  For the non-affine case, the solutions of these equations are specified 
  in terms of the positive roots $\rho_K$ \cite{CKV}, 
  where $K=1, \ldots, R_+$ with $2 R_+ + r = |G|$ and $r$ is the rank of $G$.
  In terms of the simple roots $e_i$, the positive roots can be written as
    \bea
      \rho_K 
       =     \sum_{i=1}^r n_K^i e_i,
             \label{positiveroot}
    \eea
  where $n_K^i$ are some positive integers.
  The equations of motion reduce to the following equations
    \bea
      W'_K (x) 
       =     \sum_i n_K^i W'_i (x)
       =     0.
             \label{vacuumcondition}
    \eea
  Each of these equations has $n$ roots for each positive root $\rho_K$ 
  because we have chosen every superpotential is polynomial of degree $n+1$.
  We denote these roots as $x = a_{(p, K)}$ where $p =1, \ldots, n$.
  Then, a supersymmetric vacuum is given by the value of $a_{(p, K)}$ 
  with multiplicities $M_{(p, K)} \geq 0$ which satisfy
    \bea
    N_i 
     =     \sum_{K = 1}^{R_+} \sum_{p = 1}^n M_{(p, K)} n_K^i.
    \eea
  Furthermore, the gauge group is broken as 
    \bea
    \sum_i U(N_i) 
     \rightarrow 
           \sum_K \sum_p U(M_{(p, K)}),
    \eea
  by Higgsing.
  
  For the affine case, we have one additional node in quiver diagram and the gauge group is $\sum_{i=0}^r U(N_i)$
  where $U(N_0)$ gauge group corresponds to that node.
  The classical vacua are similarly specified by the positive roots as above \cite{CKV}.
   
%%%%%%%%%%%%%%%%%%%%%%%%%%%%%%%%%%%%%%%%%%%%%%%%%%%%%%%%%%%%%%%%%%%%%%%%%%%%%%%%%%%%%%%%%%%  section 3
\section{Geometric construction}
\label{sec:geometric}
  The above gauge theories can be realized as low energy effective theories 
  on D5-branes wrapped on 2-cycles of Calabi-Yau 3-folds in the non-affine case 
  and additional D3-branes in the affine case.
  These Calabi-Yau 3-folds are constructed by non-trivially fibering the ALE spaces with $A$-$D$-$E$ singularity 
  over the complex plane.
  The important difference between the quiver gauge theory constructed above
  and the one in \cite{CKV, CFIKV} is the gauge kinetic term.
  As considered in \cite{ABSV} for $A_1$ case, the field dependent gauge kinetic term can be engineered geometrically
  by introducing the non-trivial $B$-field depending on the complex plane which is the base space of Calabi-Yau 3-fold.
  
  We consider the ALE spaces with $A$-$D$-$E$ singularity at the origin, which can be viewed 
  as the hypersurfaces $f(x, y, z) = 0$ with, e.g. for $A_{r}$ singularity,
    \bea
    f = x^2 + y^2 + z^{r+1},
    \eea
  where $x, y, z \in \textbf{C}$.
  By deforming these by relevant deformations, 
  we obtain non-singular spaces, whose defining equations are, in $A_r$ case,
    \bea
    f = x^2 + y^2 + \prod_{i = 1}^{r+1} (z + t_i),
    ~~
    \sum_{i = 1}^{r + 1} t_i = 0,
    \label{defining}
    \eea
  where $t_i$ are deformation parameters and there are $r$ independent classes of non-vanishing $S^2$'s.
  These classes intersect according to the corresponding $A$-$D$-$E$ Dynkin diagrams.
  The holomorphic volumes of $S^2$'s are defined 
  by integrals of the holomorphic 2-form $\omega = dx dy/z$ as $\alpha_i = \int_{S^2_i} \omega$.
  These $\alpha_i$ are simply related to $t_i$ in (\ref{defining}) by, in $A_r$ case,
    \bea
    \alpha_i 
     =     t_i - t_{i + 1}.
           \label{alpha}
    \eea
  For $G = D,E$ cases, the constructions are similar to the above \cite{CKV}.
  
  We consider the fibrations of these spaces over the complex plane.
  We denote the coordinate of this plane as $t$.
  If there exists D5-branes wrapping on the above $S_i^2$ and occupying the $R^{1,3}$ direction, 
  we obtain 4d non-affine $A$-$D$-$E$ quiver gauge theories as low energy effective theories on the D5-branes, 
  whose field contents are the same as those in section \ref{sec:classical}.
  If we include the D3-branes occupying $R^{1,3}$ direction, 
  the gauge theory becomes the affine quiver gauge theory \cite{DM, CKV, CFIKV}.
  Note that since $t$ plane is orthogonal to the $S^2$'s on which D5-branes wrap,
  it parametrizes the positions of the D-branes.
  Thus, $t$ corresponds to the vacuum expectation value of the adjoint chiral superfield on the D-brane world volume.
  
  In type IIB string theory, there are NSNS field $B^{NS}$ and RR field $B^R$.
  Geometrically, the complexified gauge coupling of each gauge factor of the quiver gauge theory
  corresponds to the integral over corresponding $S^2_i$ of Calabi-Yau 3-fold:
    \bea
    \left( \frac{\theta}{2 \pi} + \frac{4 \pi i}{g^2_{YM}} \right)_i
     =     \int_{S^2_i} \left( B^R + \frac{i}{g_s} B^{NS} \right).
    \eea
  Note that we have set the K\"ahler parameters to zero: $r_i \equiv \int_{S_i^2} k = 0$ where $k$ is the K\"ahler form.
  As in \cite{CKV, CFIKV}, if the background $B$-fields do not have any $t$-dependence, 
  the above quantities are constants and denote the complexified gauge coupling constants.
  On the other hand, if the background $B$-fields depend on $t$ as in \cite{ABSV}, we obtain
    \bea
    \tau_i (t)
     \equiv
           \left( \frac{\theta}{2 \pi} + \frac{4 \pi i}{g^2_{YM}} \right)_i (t)
     =     \int_{S^2_i} \left( B^R (t) + \frac{i}{g_s} B^{NS} (t) \right),
           \label{tau}
    \eea
  which produce the field-dependent gauge kinetic term of the effective theory on the D-branes,
  as in section \ref{sec:classical}.
  The point is that in order not to break the $\CN=2$ supersymmetry, 
  $B$-fields should {\it holomorphically} depend on $t$ \cite{KN}.
  Indeed, the dual IIB supergravity solution of this brane set-up, which has $\CN=2$ supersymmetry in 4d, 
  can be obtained assuming that the dilaton is constant.
  Generically, $t$-dependent $B$-fields induce the source term in the dilaton equation of motion.
  However, such a source term vanishes in the case with holomorphically $t$-dependent $B$-fields \cite{KN}.
  Therefore, the dilaton remains constant in this case and $\CN=2$ supersymmetry is not broken.
  
  The superpotentials can be turned on 
  by considering the non-trivial fibration of the ALE space over $t$-plane,
  promoting $\alpha_i$ (\ref{alpha}) to be dependent on $t$: $\alpha_i = \alpha_i (t)$.
  These $\alpha_i$ give the superpotential $W'_i (z) = \alpha_i (z)$.
  We only consider the non-monodromic fibration where $\alpha_i$ are the single-valued functions of $t$ 
  as it leads to the single trace functions $W_i (\phi_i)$ in section \ref{sec:classical}.
  Also, we choose all the degrees of the superpotentials to be $n + 1$.
  In this case, there are $n$ points in $t$ plane for each positive root $\rho_K$ (\ref{positiveroot})
  where the holomorphic volume becomes zero
    \bea
    W'_K (t)
     \equiv
           \sum_{i =1}^r n^i_K W'_i(t)
     =     0.
           \label{eom}
    \eea
  These equations correspond to the conditions for the supersymmetric vacua 
  and are same as those obtained in the gauge theory (\ref{vacuumcondition}).
  The roots of (\ref{eom}) are expressed as $t = a_{(p, K)}$ where $p = 1, \ldots n$.
  As noted above, these values correspond to the positions of D-branes and, 
  therefore, the vacuum expectation values of $\phi_i$.
  
  Therefore, we have geometrically engineered the $\CN=1$ $A$-$D$-$E$ quiver gauge theories
  with the field-dependent gauge kinetic term, which have been considered in the previous section.
  This construction is a simple generalization of $A_1$ case \cite{ABSV} to other quiver cases.
  
  Now we will comment on an important point which arises from the non-trivial fields background.
  Note that the background $B$-fields (\ref{tau}) and the positions of D-branes $a_{(p, K)}$ determine 
  the classical gauge coupling constant of each gauge factor $U(M_{(p, K)})$
    \bea
    \left( \frac{4 \pi}{g^2} \right)_{(p, K)}
     =     \sum_{i =1}^r n^i_K \Im \tau_i (a_{(p, K)}).
    \eea
  The crucial point is that, in contrast to the case in \cite{CFIKV}, these quantities could be negative
  for generic choice of the background fields and the superpotentials.
  This implies that the field theoretical description is ill-defined in that case, 
  but from the string theory point of view, the case where some of the squared gauge coupling constants are negative 
  arises from antibranes wrapping on the corresponding $S^2$'s\footnote{
    In $A_1$ case, if all the squared gauge coupling constants are negative, 
    a better field theoretical description which is supersymmetry breaking model by spurion fields \cite{LM, GMT}
    has been proposed in \cite{ABSV}.}.
  
%%%%%%%%%%%%%%%%%%%%%%%%%%%%%%%%%%%%%%%%%%%%%%%%%%%%%%%%%%%%%%%%%%%%%%%%%%%%%%%%%%%%%%%%%%%  section 4
\section{Duality in string theory}
\label{sec:weyl}
  As considered in \cite{CFIKV}, there are two types of duality in the above theories.
  The one is the geometric transition duality \cite{Vafa} 
  and the other one corresponds to Weyl reflection of $A$-$D$-$E$ groups.
  In this paper, we only consider the latter type.
  
  Weyl reflection about the simple root $e_{i_0}$ of $A$-$D$-$E$ group can be viewed as the following action
  on the simple roots:
    \bea
    e_i
     \rightarrow
           e_i - (e_i \cdot e_{i_0}) e_{i_0},
    \eea
  where the inner product of the simple roots is normalized as follows:
  $e_i \cdot e_{i_0}$ are 2 for $i = i_0$, $-1$ for $i$ connected with $i_0$ node and 0 for the other $i$.
  In the Calabi-Yau geometry, this corresponds to the change of $S^2$'s
  and leads to the following action on $\tau_i$ and the polynomial parts of the superpotential:
    \bea
    \tau_i (\phi_i)
     \rightarrow
           \tau_i (\phi_i) - (e_i \cdot e_{i_0}) \tau_{i_0} (\phi_i),
           ~~~
    W_i (\phi_i)
     \rightarrow
           W_i (\phi_i) - (e_i \cdot e_{i_0}) W_{i_0} (\phi_i).
           \label{weyl}
    \eea
  The action of the Weyl reflection on the superpotentials are exactly same as those in \cite{CFIKV}.
  But, since the gauge couplings $\tau_i$ are polynomials of $\phi_i$, 
  the higher order terms in $\phi_i$ are also affected by the Weyl reflection.
  This induces non-trivial action on the coupling constants $t_{i, k}$ in $\tau_i$.
  In fact, in the case with constant $\tau_i$, 
  this reduces to the action on the gauge coupling constants, as in \cite{CFIKV}:
  $(1/g^2)_i \rightarrow (1/g^2)_i - (e_i \cdot e_{i_0}) (1/g^2)_{i_0}$.
  
  The different looking gauge theory obtained by Weyl reflection 
  should be equivalent to the original one from the string theory perspective \cite{CFIKV}.
  This is due to ambiguities which come from the fact that 
  there is no unique way to blow up the singularity and we can determine a quiver gauge theory up to Weyl group action.
  Since the total brane charge must be conserved, 
  the ranks of the gauge groups after the transition are related with the original ranks as
    \bea
    \sum_i N'_i e'_i
     =     \sum_i N_i e_i.
    \eea
  Hence, the ranks of the gauge groups are changed under the Weyl reflection about $e_{i_0}$
  as $N'_{i_0} = N_f - N_{i_0}$ and $N'_i = N_i$ for $i\neq i_0$
  where $N_f$ is the number of flavors of $U(N_{i_0})$ gauge theory 
  when the other gauge symmetries are considered as flavor symmetries
  and $N_f \equiv \sum_{i\neq i_0} (- e_i \cdot e_{i_0}) N_i$.
  Note that the number of flavors is not changed under the Weyl reflection.
  
  As discussed in \cite{CFIKV}, this kind of duality can be considered as $\CN=1$ electric-magnetic duality 
  \cite{Seiberg, IS, Kutasov, KS, KSS, ASY} in the framework of the gauge theory.
  (See also \cite{EGK, OV, BRP, FHHU} for related approaches.)
  However, as seen above, the duality induces the non-trivial action on the field-dependent gauge kinetic term.
  This is a first example for the electric-magnetic duality in the case with the field dependent gauge kinetic term.
  We will call this as extended electric-magnetic duality.
  Fortunately, string theory has suggested that such a duality exists.
  We will check this duality field-theoretically in the subsequent sections.
  
  Before going to next, let us see the action on the superpotentials and on $\tau_i$ more explicitly.
  First of all, the action on the superpotentials (\ref{weyl}) can be written as 
    \bea
    W'_{i} (\phi_{i})
     =     \left\{ 
           \begin{array}{ll}
         - W_{i} (\phi_{i}),                    & {\rm for} ~i=i_0,\\
           W_{i} (\phi_i) + W_{i_0} (\phi_i),~~ & {\rm for} ~i ~{\rm connected ~with}~i_0, \\
           W_{i} (\phi_i),                      & {\rm for~the ~ other} ~ i.
           \end{array}
           \right.
           \label{weylsuperpot}
    \eea
  Also, for the coefficients of the gauge kinetic terms, Weyl reflection acts as
    \bea
    \tau'_{i} (\phi_{i})
     =     \left\{ 
           \begin{array}{ll}
         - \tau_{i} (\phi_{i}),                         & {\rm for}~ i=i_0,\\
           \tau_{i} (\phi_i) + \tau_{i_0} (\phi_i),  ~~ & {\rm for} ~i ~{\rm connected ~with}~i_0, \\
           \tau_{i} (\phi_i),                           & {\rm for~the ~ other} ~ i,
           \end{array}
           \right.
           \label{weylgauge}
    \eea
  Let us concentrate on the gauge theory on the $i_0$-th node.
  If we treat the gauge symmetries of the linked nodes as the weakly gauged flavor symmetries,
  we obtain $U(N_{i_0})$ gauge theory with a superpotential 
    \bea
    W
     =     \sum_{k=1}^n \frac{g_k}{k + 1} \Tr \phi^{k+1}
         + \tr \bar{Q} \phi Q + \tr m \bar{Q} Q,
           \label{superpotential}
    \eea
  where $Q$ and $\bar{Q}$ are $N_f$ fundamental and anti-fundamental superfields.
  The symbol $\tr$ denotes the trace over the flavor indices.
  These come from the bi-fundamental superfield connecting $i_0$-th node with the neighboring nodes.
  The mass term for $Q$ and $\bar{Q}$ is due to $\phi_{i'} Q_{i' i_0} Q_{i_0 i'}$ term 
  of the neighboring nodes by giving a vev of $\phi_{i'}$.
  Also, let the gauge kinetic term of this theory be
    \bea
    \Tr \tau_{i_0} (\phi) \CW^\alpha \CW_\alpha
     =     \sum_{k= 0}^n t_{k} \Tr (\phi^k \CW^\alpha \CW_\alpha),
    \eea
  where we have simplified the notation of the coupling in $\tau_{i_0}$ as $t_{i_0, k} \equiv t_k$.
  
  The Weyl action changes the gauge group to $U(N_f - N_{i_0})$ ($N_f \equiv \sum_{i\neq i_0} (- e_i \cdot e_{i_0}) N_i$)
  and the superpotential (\ref{superpotential}) to 
    \bea
    \widetilde{W}
     =   - \sum_{k = 1}^n \frac{g_k}{k + 1} \Tr \tilde{\phi}^{k + 1}
         + \sum_{k = 1}^n \frac{g_k}{k + 1} \tr m^{k + 1}
         + \tr \bar{q} \tilde{\phi} q + \tr m \bar{q} q,
           \label{dualsuperpotential}
    \eea
  where $\tilde{\phi}$ is an adjoint field of $U(N_f - N_{i_0})$ gauge group 
  and $q$ and $\bar{q}$ are the $N_f$ fundamentals and anti-fundamentals.
  The minus sign of the first term reflects 
  the Weyl action on the superpotential $W_{i_0}$ (\ref{weylsuperpot}).
  The second term comes from the Weyl action on the superpotentials of the nodes linked to $i_0$ node.
  (The trace of this term is taken over the flavor indices.)
  Furthermore, the gauge kinetic term of the dual theory becomes
    \bea
    - \sum_{k = 0}^n t_{k} \Tr (\tilde{\phi}^k \widetilde{\CW}^\alpha \widetilde{\CW}_\alpha),
    \label{dualgaugekineticterm}
    \eea
  where $\widetilde{\CW}^\alpha$ is the field strength superfield of the dual theory.
  
  In the theory corresponding to a node connected with $i_0$-th node, 
  as noted above, the dual superpotential of this theory contributes to the second term of (\ref{dualsuperpotential})
  because the Weyl reflection induces an additional term $W_{i_0} (\tilde{\phi}_i)$ (\ref{weylsuperpot})
  where $\tilde{\phi}_i$ is an chiral superfield of this dual theory.
  On the other hand, the gauge kinetic term is affected as follows:
    \bea
    \tau_i (\phi_i)
     \rightarrow
           \tau_i (\tilde{\phi}_i) + \tau_{i_0} (\tilde{\phi}_i)
     =     \tau_i (\tilde{\phi}_i) + \sum_{k = 0}^n t_k \tilde{\phi}_i^k,
    \eea
  as easily extracted from (\ref{weylgauge}).

%%%%%%%%%%%%%%%%%%%%%%%%%%%%%%%%%%%%%%%%%%%%%%%%%%%%%%%%%%%%%%%%%%%%%%%%%%%%%%%%%%%%%%%%%%%%%%%%%%%%  section 5
\section{Extended electric-magnetic duality}
\label{sec:fieldtheory}
  We have seen that the string theory construction has suggested an extension of $\CN=1$ electric-magnetic duality
  to the case where the gauge kinetic term depends on the adjoint chiral superfields.
  In what follows, we concentrate on a particular node of the quiver and consider the duality 
  from field-theoretical point of view.

  Let us specify the model.
  Consider $\CN=1$, $U(N_c)$ gauge theory with an adjoint chiral superfield $\phi$ 
  and $N_f$ fundamental and anti-fundamental superfields $Q$ and $\bar{Q}$, 
  and also with a gauge kinetic term which depends on the adjoint chiral superfield:
    \bea
    \int d^2 \theta \Tr \tau (\phi) \CW^\alpha \CW_\alpha,
    ~~~
    \tau (\phi)
     =     \sum_{k=0}^{m} t_k \phi^k,
    \eea
  and a superpotential (\ref{superpotential})
    \bea
    W
     =     \sum_{k=1}^n \frac{g_k}{k + 1} \Tr \phi^{k+1}
         + \tr \bar{Q} \phi Q + \tr m \bar{Q} Q.
           \label{fullsuperpotential}
    \eea
  Without loss of the generality, the mass matrix $m$ can be chosen to be diagonal.
  We will use the indices $f = 1, \ldots, N_f$ to label the flavors.
  In this notation, the diagonal components of the mass matrix are written as $m_f$.

  The vacua of this theory can be divided into two types: confining and Higgs vacua.
  Classically, the confining vacua correspond to 
  the vacuum expectation values such that 
  $\langle Q \rangle = \langle \bar{Q} \rangle = 0$ 
  and 
    \bea
    \langle \phi \rangle = \diag (a_1, a_2, \ldots, a_N),
    \label{vevphi}
    \eea 
  where $a_i$ are determined from the solutions of the F-term equation: 
    \bea
    W'(x)
     \equiv
           g_n \prod_{i = 1}^n (x - a_i)
     =     0.
           \label{F-term}
    \eea
  Note that the other terms contributing to the F-term equation vanish in these vacua.
  Indeed, as we have seen in the quiver case, 
  the gauge kinetic term, $\tau(\phi) \CW^\alpha \CW_\alpha$, produces an additional term in the F-term equation
  such as $\partial_\phi \tau (\phi) \lambda^\alpha \lambda_\alpha$ where $\lambda^\alpha$ is the gluino, 
  but this term vanishes because we are interested in the vacua where the vacuum expectation values of the fermions are zero.
  
  The Higgs vacua correspond to the case 
  where some of the diagonal elements of $\langle \phi \rangle$ are equal to the mass parameters
  and $Q$ and $\bar{Q}$ have non-zero vacuum expectation values which are determined from the F-term equation:
    \bea
    (W'(\phi))_{ij} + \sum_f Q_{j}^f \bar{Q}_{i f} 
     =     0,
           \label{F-termHiggs}
    \eea
  where $i, j = 1, \ldots, N$ are the gauge indices.
  As above, the gauge kinetic term does not contribute to the classical equation (\ref{F-termHiggs}).
  
  In subsection \ref{subsec:singletrace}, we begin to consider the case without $\bar{Q} \phi Q$ and $m \bar{Q}Q$ terms.
  In this case, the flavors are massless and, after integrating out the adjoint fields,
  the theory splits into a set of the decoupled SQCD theories with the massless flavors.
  Therefore, the stable vacua exist if \cite{KS}
    \bea
    \frac{N_c}{n} \leq N_f.
    \label{stable}
    \eea
  We will see the dual description of the above theory, after reviewing the constant $\tau$ case.
  Then, we will turn to the case with full superpotential (\ref{fullsuperpotential}) in subsection \ref{subsec:full}.
  
%%%%%%%%%%%%%%%%%%%%%%%%%%%%%%%%%%%%%%%%%%%%%%%%%%%%%%%%%%%   subsection 5.1
\subsection{Single trace superpotential case}
\label{subsec:singletrace}
  We consider the case where the superpotential is 
    \bea
    W
     =     \sum_{k = 1}^n \frac{g_k}{k + 1} \Tr \phi^{k + 1}.
           \label{electricsuperpotential}
    \eea
  We first review the case where $\tau$ is constant, i.e. $t_k=0$ for $k > 0$.
  In this case, the dual description of the theory has been obtained in \cite{Kutasov, KS, KSS}, 
  which is $U(\tilde{N}_c)$ ($\tilde{N}_c = n N_f - N_c$) gauge theory with $N_f$ fundamental and anti-fundamental superfields $q$ and $\bar{q}$,
  gauge singlet superfields $M_i$ ($i = 1, \ldots, n$) and an adjoint chiral superfield $\tilde{\phi}$.
  The singlet fields $M_i$ are identified with the meson superfields in the original theory as
    \bea
    M_i 
     =     \bar{Q} \phi^{i - 1} Q, 
           ~~~
           i = 1, \ldots, n.
           \label{matching}
    \eea
  It is not necessary to introduce the other meson fields corresponding to $\bar{Q} \phi^\ell Q$ ($\ell > n$), 
  since such fields can be eliminated by the chiral ring relation.
  In addition, the superpotential of the dual theory is \cite{KSS}
    \bea
    \widetilde{W}
     =   - \sum_{k=1}^n \frac{g_k}{k + 1} \Tr \tilde{\phi}^{k + 1} 
         + \frac{1}{\mu^2} \sum_{k=1}^{n} g_k \sum_{i = 1}^k M_i \bar{q} \tilde{\phi}^{k - i} q,
           \label{magneticsuperpotential}
    \eea
  where a parameter $\mu$ has been introduced in order for the dimension of the second term to be correct.
  This duality is a generalization of the electric-magnetic duality (Seiberg duality) in $\CN=1$ SQCD \cite{Seiberg}
  to the case with an adjoint chiral superfield and a tree level single trace superpotential.
  Below we refer to the original and dual theories as electric and magnetic theories respectively.
  
  The global symmetries of both theories are the same:
  there is $SU(N_f) \times SU(N_f) \times U(1)_R \times U(1)_J$ symmetry.
  The $U(1)$ charges of the fields (and the parameters) of the electric theory are in Table \ref{tab1},
    \begin{table}[t]
    \begin{center}
    \begin{tabular}{c|c|c|c|c|c|c|c}
             & $\phi$ & $\CW^\alpha$ & $Q$ (or $\bar{Q}$) & $g_k$   & $\Lambda^{2 N_c - N_f}$ \\ \hline
    $U(1)_R$ & $2$    & $1$          & $0$               & $- 2 k$ & $2 (2 N_c - N_f)$       \\ \hline
    $U(1)_J$ & $0$    & $1$          & $1$               & $2$     & $0$                     \\
    \end{tabular}
    \caption{$U(1)$ charges of the electric fields and the parameters.}
    \label{tab1}
    \end{center}
    \end{table}
  where $\Lambda$ is the dynamical scale of the electric theory
  and the superspace coordinate $\theta$ also has charge 1.
  Note that we have allowed the coupling constants $g_k$ and $t_k$ transform non-trivially,
  as $U(1)_R$ and $U(1)_J$ become the symmetries with the superpotential and the gauge kinetic term. 
  Also, the charges of the fields of the magnetic theory are in Table \ref{tab2},
    \begin{table}[t]
    \begin{center}
    \begin{tabular}{c|c|c|c|c|c|c|c|c}
                      & $\tilde{\phi}$ & {\small $\widetilde{\CW^\alpha}$} & $q$ (or $\bar{q}$)                               & $M_\ell$                & $g_k$   & $\mu$                                                                  & $\tilde{\Lambda}^{2 \tilde{N}_c - N_f}$                                                         \\ \hline
    {\small $U(1)_R$} & $2$            & $1$                               & $\frac{(n - 1)(N_c - \tilde{N}_c)}{\tilde{N}_c}$ & {\small $2 (\ell - 1)$} & $- 2 k$ & {\footnotesize $-2+$}$ \frac{(n - 1)(N_c - \tilde{N}_c)}{\tilde{N}_c}$ & {\footnotesize $4 \tilde{N_c} - 2 N_f +$}$\frac{2 N_f (n - 1)(N_c - \tilde{N}_c)}{\tilde{N}_c}$ \\ \hline
    {\small $U(1)_J$} & $0$            & $1$                               & $\frac{N_c}{\tilde{N}_c}$                        & $2$                     & $2$     & $1 + \frac{N_c}{\tilde{N}_c}$                                          & $\frac{2 N_f (N_c - \tilde{N}_c)}{\tilde{N}_c}$                                                 \\
    \end{tabular}
    \caption{$U(1)$ charges of the magnetic fields and the parameters.}
    \label{tab2}
    \end{center}
    \end{table}
  where $\tilde{\Lambda}$ is the dynamical scale of the magnetic theory.
  A non-trivial check of this duality is to compare the 't Hooft anomalies of the theories. 
  It has been shown that they perfectly match in the case with the truncated superpotential \cite{KS}.
  
  An important ingredient of $\CN=1$ duality is the matching relation of the dynamical scales.
  In the case here, the relation is
    \bea
    \Lambda^{2 N_c - N_f} \tilde{\Lambda}^{2 \tilde{N}_c - N_f}
     =     g_n^{- 2 N_f} \mu^{2 N_f}.
    \eea
  One can easily check that this relation is consistent with the above charge assignment.
  Also, in \cite{KSS}, it has been shown that this is consistent with the deformations of the theory 
  by the mass terms of the flavors.
  
  It is worth noting that, on general grounds,
  the coefficients of $\Tr \tilde{\phi}^{k + 1}$ and $M_i \bar{q} \tilde{\phi}^{k - i} q$ 
  in the magnetic superpotential are generic functions of $g_k$.
  However, we can fix these coefficients as in (\ref{magneticsuperpotential}).
  First of all, in the electric theory, according to (\ref{vevphi}), the gauge symmetry is broken to $\prod_{i = 1}^n U(r_i)$ where $\sum_i r_i = N_c$.
  ($r_i$ denote the number of the eigenvalues of $\langle \phi \rangle$ which are equal to $a_i$.)
  Supposing that the underlying $U(N_c)$ gauge theory is weakly coupled at the mass scale 
  which is specified by the above superpotential, 
  the theory splits in the low energy into a set of decoupled SQCD theories 
  with $U(r_i)$ gauge groups and $N_f$ flavors
    \footnote{This is the case where all the roots of (\ref{F-term}) are different from each other.
              In the case where some of $a_i$ coincide, i.e. $W' = \prod_{i = 1}^r (x - a_i)^{n_i}$ ($r<n$),
              each decoupled theory has a superpotential as $\Tr \phi_i^{n_i}$.}.
  In the dual theory, the coefficients of $\Tr \tilde{\phi}^{k + 1}$ have been fixed 
  such that the magnetic superpotential has the same critical points $a_i$ as those in the electric theory.
  Then, we observe a similar gauge symmetry breaking pattern: 
  $U(n N_f - N_c) \rightarrow \prod_i U(\tilde{r}_i)$.
  The claim is that $\tilde{r}_i = N_f - r_i$,
    \footnote{In the case corresponding to the above footnote, 
              the corresponding gauge group is $U(n_i N_f - r_i)$.}
  in order to obtain one-to-one correspondence 
  between each $U(\tilde{r}_i)$ SQCD theory and each of the decoupled SQCD theories in the electric theory
  under Seiberg duality \cite{Seiberg}
    \footnote{In the case corresponding to the above footnotes,
              we demand that each decoupled theory in the magnetic theory is related with 
              each decoupled theory in the electric theory by Kutasov duality \cite{Kutasov, KS}
              with the truncated superpotential}.
  Also, the magnetic theory should split to $U(N_f - r_i)$ SQCD theories with $N_f$ flavors and mesons, 
  as the electric theory does.
  This determines the coefficients of $M_i \bar{q} \tilde{\phi}^{k - i} q$ 
  and leads to (\ref{magneticsuperpotential}) \cite{KSS}.
  
%%%%%%%%%%%%%%%%%%%%%%%%%%%%%%%%%%%%%%%%%%%%%%%%%%%%%%%%%%%
\subsubsection*{Magnetic gauge kinetic term}
  We now turn to the analysis of the gauge kinetic term.
  We first note that inclusion of the $\phi$-dependent part of $\tau$ 
  does not change the structure of the classical chiral ring.
  The classical chiral ring relations, i.e. a set of constraints on the gauge invariant operators
  follows from the F-term equation 
  (and a constraint on characteristic polynomial: $f(\phi)=0$ with $f(x) = \det (x - \phi)$).
  Indeed, as we have seen above, 
  $\tau(\phi) \CW^\alpha \CW_\alpha$ term does not affect the classical solution.
  On the other hand, the quantum chiral ring is modified by the existence of the field dependent part of $\tau$
  because the gluino confines in the confining vacua and leads to the non-zero vacuum expectation value of 
  $\langle \lambda^\alpha \lambda_\alpha \rangle$.
  We will see this in next section by analyzing the generalized Konishi anomaly equations.
  
  Now, consider the magnetic superpotential.
  In general, it could depend on $t_k$ as well as $g_k$.
  Recall however that the magnetic superpotential has been determined 
  such that it has the same critical points as those of the electric theory and 
  it is consistent with the decoupling of the SQCD theories in the magnetic theory.
  This process can be applied to the case with the field-dependent gauge kinetic term:
  if the magnetic superpotential depends on $t_k$, 
  we can no longer obtain the same critical points.
  Also, $t_k$-dependent $M_i \bar{q} \tilde{\phi}^{k - i} q$ terms obviously make decoupled SQCD theories
  to couple each other.
  Therefore, the magnetic superpotential cannot depend on $t_k$.
  
  On the other hand, the gauge kinetic term of the magnetic theory can be written generally as
    \bea
    \sum_{k=0}^n \tilde{t}_k \Tr \tilde{\phi}^k \widetilde{\CW}^\alpha \widetilde{\CW}_\alpha,
    \eea
  where $\tilde{t}_k$ are some functions of the parameters in the electric theory,
  which relate the coupling constants of the electric theory with those of the magnetic theory.
  As the coupling constants in the magnetic superpotential have been fixed 
  such that the magnetic theory correctly behaves as the dual of the original one, 
  we have to choose the correct form of the functions $\tilde{t}_k (t, g)$.
  We will see below that this is simply
    \bea
    \tilde{t}_k
     =   - t_k,
           \label{tildet}
    \eea
  for $k = 1, \ldots, m$.
  The lowest coupling constant, i.e. $\tilde{t}_0$, can also be determined 
  from the matching relation of the dynamical scales (\ref{matchingrelation}).
  
  Let us see (\ref{tildet}) is indeed the case.
  We first consider the matching relation of the dynamical scales of the electric and magnetic theories. 
  As we have seen above, the matching relation in the case with constant $\tau$ is (\ref{matching})
    \bea
    \Lambda^{2 N_c - N_f} \tilde{\Lambda}^{2 \tilde{N}_c - N_f}
     =     g_n^{- 2 N_f} \mu^{2 N_f}.
           \label{matchingrelation}
    \eea
  In the case with $\tau(\phi)$,
  we can assign $U(1)_R$ and $U(1)_J$ charges to $t_k$ and $\tilde{t}_k$ as
    \begin{center}
    \begin{tabular}{c|c|c}
             & $t_k$   & $\tilde{t}_k$ \\ \hline
    $U(1)_R$ & $- 2 k$ & $- 2 k$       \\ \hline
    $U(1)_J$ & $0$     & $0$           \\
    \end{tabular}
    \end{center}
  in addition to the charge assignment in table \ref{tab1} and \ref{tab2}.
  It follows from the above global $U(1)$ charges and also the consistency 
  with the mass (of the flavor) deformation as in \cite{KSS}
  that this relation cannot change even if we add the parameters $t_k$ and $\tilde{t}_k$ to the theory.
  Therefore, the relation is valid in the case we consider here.
  
  By integrating the massive vector superfields and the massive adjoint field out in both theories, 
  the matching relation leads to 
    \bea
    \Lambda_i^{3 r_i - N_f} \tilde{\Lambda}_i^{3 \tilde{r}_i - N_f}
     =     (-)^{N_f - r_i} g_n^{- N_f} \mu^{2 N_f} 
           e^{- 2 \pi i (T(a_i) + \tilde{T}(a_i))}
           \prod_{j \neq i} (a_i - a_j)^{- N_f},
           \label{relation}
    \eea
  for each $i$.
  We have defined as $T(x) = \tau(x) - t_0$ and $\tilde{T}(x) = \tilde{\tau}(x) - \tilde{t}_0$.
  $\Lambda_i$ and $\tilde{\Lambda}_i$ are dynamical scales of $U(r_i)$ and $U(\tilde{r}_i)$ theories 
  ($\tilde{r}_i \equiv N_f - r_i$), 
  which are defined by the matching of the gauge coupling constants:
    \bea
    \Lambda^{2 N_c - N_f}
     =     \Lambda_i^{3 r_i - N_f} \frac{e^{2 \pi i T(a_i)}}{(W''(a_i))^{r_i}} 
           \prod_{j \neq i} (a_i - a_j)^{2 r_j},
           \label{matchingg}
    \eea
  and the similar equations for the magnetic variables.
  In (\ref{matchingg}), $\prod_{j \neq i} (a_i - a_j)^{2 r_j}$ factor comes 
  from the integration of the massive vector superfields and
  $(W''(a_i))^{r_i}$ factor is due to the massive adjoint field.
  Furthermore, we add the factor $e^{2 \pi i T(a_i)}$ because the gauge kinetic term depends on the adjoint field.
  
  Finally, we note that the relation (\ref{relation}) should be consistent with the decoupling 
  of the SQCD theories in the electric and magnetic theories at low energy.
  This implies that the following relations
    \bea
    \Lambda_i^{3 r_i - N_f} \tilde{\Lambda}_i^{3 \tilde{r}_i - N_f}
     =     (-)^{N_f - r_i} \mu_i^{N_f}
    \eea
  are satisfied for each decoupled SQCD \cite{Seiberg}, 
  where $\mu_i$ are the parameters in the magnetic superpotentials of $U(\tilde{r}_i)$ SQCD theories, 
  $\mu_i^{-1} \bar{q}_i q_i M_i$.
  Since we can show that $g_n^{- 1} \mu^{2} \prod_{i \neq j} (a_i - a_j)^{- 1} = \mu_i$ as in \cite{KSS}, 
  we therefore obtain 
    \bea
    \sum_{k = 1}^{m} t_k (a_i)^k 
     =  - \sum_{k = 1}^{m} \tilde{t}_k (a_i)^k, 
          \label{vev}
    \eea
  which implies $\tilde{t}_k = - t_k$ for $k = 1, \ldots, m$. 
  In principle, (\ref{vev}) could have an additional integer term.
  However, such a term must vanish since there is no way to satisfy the equality with that term.
  Note that the parameters $t_k$ are the values at the energy scale where the gauge symmetry is broken.
  
  Note also that the argument above is valid only in the region where the gauge coupling constant is small.
  We will see in section \ref{subsec:R} that (\ref{tildet}) can be verified by using a different method.
  
%%%%%%%%%%%%%%%%%%%%%%%%%%%%%%%%%%%%%%%%%%%%%%%%%%%%%%%%%%%  subsection 5.2
\subsection{Generic superpotential case}
\label{subsec:full}
  Based on the above argument, let us consider the case with more generic superpotential 
  which has been appeared in the string theory construction:
    \bea
    W
     =     \sum_{k = 1}^n \frac{g_k}{k + 1} \Tr \phi^{k + 1}
         + \tr \bar{Q} \phi Q + \tr m \bar{Q} Q.
           \label{superpotential2}
    \eea
  As discussed in \cite{ASY, EGKRS}, by flowing from the theory considered in the previous subsection or in \cite{KSS}, 
  we can deduce that the dual superpotential becomes
    \bea
    \widetilde{W}
     =   - \sum_{k=1}^n \frac{g_k}{k + 1} \Tr \tilde{\phi}^{k + 1} 
         + \frac{1}{\mu^2} \sum_{k=1}^{n} g_k \sum_{i = 1}^k M_i \bar{q} \tilde{\phi}^{k - i} q
         + \lambda M_2 + m M_1.
           \label{magneticsuperpotential3}
    \eea
  By the relations (\ref{magneticsuperpotential}), 
  the last two terms correspond $\bar{Q} \phi Q$ and the mass deformations.
  What we have to check about this superpotential is whether the deformation terms do not spoil the separation
  of the SQCD theories or not, as we have discussed in the previous subsection.
  But it is obviously trivial 
  since the last two terms have no room to mix the operators of the different gauge factors.
  
  The analysis of the dual gauge kinetic term is the same as that of the previous subsection
  and we do not repeat here.
  The conclusion is $\tilde{t}_k = - t_k$.
  This is exactly same as what has been expected in the string theory (\ref{dualgaugekineticterm}).
  
  While we have formulated a magnetic dual, 
  the magnetic superpotential (\ref{magneticsuperpotential3}) is different from 
  the one expected from the string theory duality (\ref{dualsuperpotential}).
  In fact, the dual theory obtained by Weyl reflection in string theory does not include the meson fields
  and the gauge groups are also different: $U(N_f - N_c)$ in the stringy dual theory,
  and $U(n N_f - N_c)$ in the magnetic theory in present section.
  However, one can show that the magnetic theory reduces to the stringy dual one 
  after integrating out the mesons and (anti-)fundamentals
  and Higgsing to $U(N_f - N_c)$ gauge theory as in \cite{CFIKV}.
  As we have already seen, the gauge kinetic term which depends on the adjoint chiral superfield does not affect 
  the classical equations of motion.
  Therefore, the discussion is the same as that in the theory with constant $\tau$.
  
%%%%%%%%%%%%%%%%%%%%%%%%%%%%%%%%%%%%%%%%%%%%%%%%%%%%%%%%%%%%%%%%%%%%%%%%%%%%%%%%%%%%%%%%%%%%%%%%%%%%  section 6
\section{Duality map of the chiral operators}
\label{sec:dualitymap}
  In this section, let us consider the duality map between the chiral operators in the electric theory 
  and the magnetic ones.
  First of all, we consider the operators $\Tr \CW^\alpha \CW_\alpha$ 
  (and $\Tr \widetilde{\CW}^\alpha \widetilde{\CW}_\alpha$).
  As already seen above, the matching relations of the dynamical scales of the decoupled SQCD theories are
    \bea
    \Lambda_i^{3 r_i - N_f} \tilde{\Lambda}_i^{3 \tilde{r}_i - N_f}
     =     (-)^{N_f - r_i} \mu_i^{N_f},
    \eea
  for each $i$.
  In each $U(r_i)$ SQCD theory,
  the gauge coupling constant receives one-loop correction and the gauge kinetic term is renormalized as
  $(3 r_i - N_f) \log (\Lambda_i/M) \Tr \CW^{i \alpha} \CW^i_\alpha$ in the electric theory and 
  $(3 \tilde{r}_i - N_f) \log (\tilde{\Lambda}_i/M) \Tr \widetilde{\CW}^{i \alpha} \widetilde{\CW}^i_\alpha$ 
  in the magnetic theory.
  If we take a derivative with respect to $\log \Lambda_i$ and use (\ref{relation}) as in \cite{IS, KSS}, 
  we obtain the following relations:
    \bea
    {\rm Tr}_{U(r_i)} \CW^{i \alpha} \CW^i_\alpha
     =   - {\rm Tr}_{U(\tilde{r}_i)} \widetilde{\CW}^{i \alpha} \widetilde{\CW}^i_\alpha,
           \label{WW=-WW}
    \eea
  for each $i$.
  These imply that the gauge coupling constant of each decoupled SQCD theory in the electric theory is
  different by sign from the magnetic one. 
  
  To check the other relations in terms of more complicated operators, 
  it is convenient to use the generalized Konishi anomaly equations, 
  as in \cite{Mazzucato} for the constant $\tau$ case.
  Thus, we first derive these equations in subsection \ref{subsec:GKA}.
  Then, we will consider the duality map of the operators in subsection \ref{subsec:R} and \ref{subsec:T}.

%%%%%%%%%%%%%%%%%%%%%%%%%%%%%%%%%%%%%%%%%%%%%%%%%%  subsection 6.1
\subsection{Generalized Konishi anomaly equations}
\label{subsec:GKA}
  Let us derive the generalized Konishi anomaly equations 
  in the electric and magnetic theories.
  We define the generating functions of the one-point functions in the electric theory as
    \bea
    R(z)
    &=&  - \frac{1}{64 \pi^2} \left< \Tr \frac{\CW^\alpha \CW_\alpha}{z - \phi} \right>,
           \nonumber \\
    T(z)
    &=&    \left< \Tr \frac{1}{z - \phi} \right>,
           \nonumber \\
    M(z)_f^{f'}
    &=&    \left< \bar{Q}_f \frac{1}{z - \phi} Q^{f'} \right>,
    \eea
  where we have ignored the fermionic one-point function.
  The generalized Konishi anomaly equations in terms of these variables are\footnote{
    These anomaly equations were derived also in \cite{Ferrarif} recently.}
  :
    \bea
    R(z)^2
    &=&    \Bigg[ W'(z) R(z) \Bigg]_-,
           \nonumber \\
    2 R(z) T(z)
    &=&    \Bigg[ W'(z) T(z) \Bigg]_- + 32 \pi^2 i \Bigg[ \tau'(z) R(z) \Bigg]_- + M(z),
           \nonumber \\
    - \delta^{f'}_{f} R(z)
    &=&    \Bigg[ M(z)_f^{f'} (z + m_{f'}) \Bigg]_-,
           \label{GKAelectric}
    \eea
  which can be obtained by generalizing the arguments in \cite{CDSW, Seiberg2, CSW2, IM, Ferrari}.
  In the last equation, the flavor index $f'$ is not contracted.
  
  In the magnetic theory, we can also define
    \bea
    \widetilde{R}(z)
    &=&  - \frac{1}{64 \pi^2} \left< \Tr \frac{\widetilde{\CW}^\alpha \widetilde{\CW}_\alpha}{z - \tilde{\phi}} \right>,
           \nonumber \\
    \widetilde{T}(z)
    &=&    \left< \Tr \frac{1}{z - \tilde{\phi}} \right>,
           \nonumber \\
    \widetilde{M}(z)_f^{f'}
    &=&    \left< \bar{q}_f \frac{1}{z - \tilde{\phi}} q^{f'} \right>.
    \eea
  In terms of these, the anomaly equations can be obtained as
    \bea
    \widetilde{R}(z)^2
    &=&  - \Bigg[ W'(z) \widetilde{R}(z) \Bigg]_-,
           \nonumber \\
    2 \widetilde{R}(z) \widetilde{T}(z)
    &=&  - \Bigg[ W'(z) \widetilde{T}(z) \Bigg]_- - 32 \pi^2 i \Bigg[ \tau'(z) \widetilde{R}(z) \Bigg]_- 
         + \Bigg[ \widetilde{M}(z) A'(z) \Bigg]_-,
           \nonumber \\
    - \delta^{f'}_{f} \widetilde{R}(z)
    &=&    \Bigg[ \widetilde{M}(z)_f^{f''} A(z)_{f''}^{f'} \Bigg]_-,
           \label{GKAmagnetic}
    \eea
  where
    \bea
    A(z)
     =     \frac{1}{\mu^2} \sum_{k = 1}^n g_k \sum_{i = 1}^k M_i z^{k - i}.
    \eea
  Note that $M_i \bar{q} \phi^{k - i} q$ terms in the magnetic superpotential do not contribute 
  to the first equation of (\ref{GKAmagnetic})
  because the terms with $\bar{q} \CW^\alpha$ and $\CW^\alpha q$ are zero in the chiral ring.
  
  Another important point of these anomaly equations (\ref{GKAelectric}) and (\ref{GKAmagnetic}) is that 
  the $\phi$(or $\tilde{\phi}$)-dependence of the gauge kinetic term does not affect 
  the anomaly equation for $R(z)$ (or $\widetilde{R}(z)$), as noted in \cite{IM, Ferrari}.
  In other words, $t_k$ and $\tilde{t}_k$ do not enter in those equations.
  This is crucial in the analysis in subsequent subsections.

%%%%%%%%%%%%%%%%%%%%%%%%%%%%%%%%%%%%%%%%%%%%%%%%%%  subsection 6.2
\subsection{Duality map of $\Tr \phi^k \CW^\alpha \CW_\alpha$ operators}
\label{subsec:R}
  In this subsection, we consider the operators $\Tr \phi^k \CW^\alpha \CW^\alpha$
  ($\Tr \tilde{\phi}^k \widetilde{\CW}^\alpha \widetilde{\CW}^\alpha$ in the magnetic theory).
  We expect from the argument in previous section that the following duality map of the operators are satisfied:
    \bea
    \Tr \phi^k \CW^\alpha \CW_\alpha
     =   - \Tr \tilde{\phi}^k \widetilde{\CW}^\alpha \widetilde{\CW}_\alpha.
           \label{operatormatching}
    \eea
  We will check this relation in the vacuum.
  It should be noted that in the case without $\bar{Q} \phi Q$ and $m \bar{Q}Q$ terms, 
  the argument in the rest of this section might be invalid.
  More precisely, we obtain $\langle \Tr \phi^k \CW^\alpha \CW_\alpha \rangle = 0$, 
  as we can see from the anomaly equation for $M(z)$.
  This is because the flavors remain massless at IR.
  Therefore, we will consider the full superpotential (\ref{fullsuperpotential}) below.
  
  Since $\langle \Tr \phi^k \CW^\alpha \CW_\alpha \rangle = 0$ in the classical vacuum,
  the classical analysis cannot be non-trivial check of the duality map (\ref{operatormatching}).
  However, they could have non-zero expectation values in the quantum vacuum, as can be seen from the anomaly equations.
  Indeed, we can relate $R(z)$ to $\widetilde{R}(z)$ 
  by using the generalized Konishi anomaly equations and this will be a non-trivial check of (\ref{operatormatching}).
  Let us see this below.
  
  The generalized Konishi anomaly equations (\ref{GKAelectric}) and (\ref{GKAmagnetic}) for $R(z)$ and $\widetilde{R}(z)$,
  can be rewritten as
    \bea
    R(z)^2
     =     W'(z) R(z) + \frac{f(z)}{4},
           ~~~
    \widetilde{R}(z)^2
     =   - W'(z) \widetilde{R}(z) + \frac{\tilde{f}(z)}{4},
    \eea
  where $f(z)$ and $\tilde{f}(z)$ are the polynomials of degree $n - 1$.
  These equations can be easily solved as
    \bea
    R(z)
     =     \frac{1}{2} \left( W'(z) - \sqrt{W'(z)^2 + f(z)} \right),
           ~~~
    \widetilde{R}(z)
     =     \frac{1}{2} \left( - W'(z) + \sqrt{W'(z)^2 + \tilde{f}(z)} \right),
    \eea
  where the signs of the square roots have chosen to be consistent 
  with the large $z$ behavior of $R(z)$ and $\widetilde{R}(z)$.
  From the above forms, we can see that $R(z)$ and $\widetilde{R}(z)$ have cuts in the complex $z$ plane
  and are, respectively, meromorphic functions on Riemann surfaces $\Sigma$ and $\tilde{\Sigma}$ of genus $n - 1$:
  $y^2 = W'(z)^2 + f(z)$ and $\tilde{y}^2 = W'(z)^2 + \tilde{f}(z)$.
  Let us denote by $\alpha_i$ and $\tilde{\alpha}_i$ $\alpha$-cycles of $\Sigma$ and $\tilde{\Sigma}$ respectively.
  
  The polynomials $f(z)$ and $\tilde{f}(z)$ are completely fixed \cite{CDSW} by 
    \bea
    - \frac{1}{64 \pi^2} \langle {\rm Tr}_{U(r_i)} \CW^{i \alpha} \CW^i_\alpha \rangle
    &=&    \frac{1}{2 \pi i} \oint_{\alpha_i} R(z) dz,
           \nonumber \\
    - \frac{1}{64 \pi^2} \langle {\rm Tr}_{U(\tilde{r}_i)} \widetilde{\CW}^{i \alpha} \widetilde{\CW}^i_\alpha \rangle
    &=&    \frac{1}{2 \pi i} \oint_{\tilde{\alpha}_i} \widetilde{R}(z) dz.
    \eea
  It follows from these equations and (\ref{WW=-WW}) that $f(z) = \tilde{f}(z)$.
  Therefore, we obtain $R(z) = - \widetilde{R}(z)$, which implies
    \bea
    \langle \Tr \phi^k \CW^\alpha \CW_\alpha \rangle
     =   - \langle \Tr \tilde{\phi}^k \widetilde{\CW}^\alpha \widetilde{\CW}_\alpha \rangle.
           \label{relation2}
    \eea
  
  Note that this could be an alternative check of the magnetic gauge kinetic term.
  Indeed, as we have noted above, $M_i \bar{q} \phi^{k - 1} q$ terms and the magnetic gauge kinetic term
  in the Lagrangian do not contribute to the anomaly equation for $\widetilde{R}(z)$.
  What we have assumed in the above argument is 
  that the polynomial part of the magnetic superpotential is $- W(\tilde{\phi})$ and
  the relations (\ref{WW=-WW}).
  However, these follows from that the electric and magnetic superpotentials have the same critical points and 
  that both theories split into the decoupled SQCD theories at low energy.
  Once we have derived (\ref{relation2}), 
  we then obtain the following relations by taking derivatives of the partition functions 
  with respect to $t_k$ and $\tilde{t}_k$:
    \bea
    \frac{\partial Z}{\partial t_k}
     \sim  \langle \Tr \phi^k \CW^\alpha \CW_\alpha \rangle
     =   - \langle \Tr \tilde{\phi}^k \widetilde{\CW}^\alpha \widetilde{\CW}_\alpha \rangle
     \sim- \frac{\partial \widetilde{Z}}{\partial \tilde{t}_k}
    \eea
  Since the duality implies $Z = \widetilde{Z}$ at least for the holomorphic sector, 
  thus we can conclude that the magnetic gauge kinetic term is (\ref{tildet}).
  
%%%%%%%%%%%%%%%%%%%%%%%%%%%%%%%%%%%%%%%%%%%%%%%%%%  subsection 6.3
\subsection{Duality map of $\Tr \phi^k$ operators}
\label{subsec:T}
  Finally, we analyze the operator relations between $\Tr \phi^k$ and $\Tr \tilde{\phi}^k$.
  In the theory with constant $\tau$, it has been known \cite{KSS} that the duality map can be written as
    \bea
    \Tr \phi^k
     = - \Tr \tilde{\phi}^k 
       + \frac{k}{\mu^2} \sum_{i = 1}^{k - 1} M_i \bar{q} \tilde{\phi}^{k - 1 - i} q + \ldots, 
         \label{matchingT}
    \eea
  where ellipsis denotes the constant term.
  In \cite{KSS}, these relations have been checked by substituting the classical vacuum expectation values.
  Also, they have been analyzed in \cite{Mazzucato} by using the generalized Konishi anomaly equations.
  On general grounds, we can expect that these relations can be deformed 
  by the term with the operators $\Tr \tilde{\phi}^k \widetilde{\CW}^\alpha \widetilde{\CW}_\alpha$ 
  and the terms involving $t_k$, in the case with $\tau(\phi)$.
  Let us show below that such terms do not exist by making use of the generalized Konishi anomaly equations.
  
  The third equations of the generalized Konishi anomaly equations (\ref{GKAelectric}) and (\ref{GKAmagnetic}):
    \bea
    - \delta^f_{f'} R(z)
     =     \Bigg[ M(z)_{f'}^{f} (z + m_f) \Bigg]_-,
           ~~~
    - \delta^f_{f'} \widetilde{R}(z)
     =     \Bigg[ \widetilde{M}(z)_{f'}^{f} A(z) \Bigg]_-
    \eea
  imply that the $t_k$-dependence cannot enter in $M(z)$ and $\widetilde{M}(z)$, 
  since $R(z)$ and $\widetilde{R}(z)$ are independent of $t_k$.
  On the other hand, the second equations of (\ref{GKAelectric}) and (\ref{GKAmagnetic}) are
    \bea
    2 R(z) T(z)
    &=&    \Bigg[ W'(z) T(z) \Bigg]_- + 32 \pi^2 i \Bigg[ \tau'(z) R(z) \Bigg]_- + M(z),
           \nonumber \\
    - 2 R(z) \widetilde{T}(z)
    &=&  - \Bigg[ W'(z) \widetilde{T}(z) \Bigg]_- + 32 \pi^2 i \Bigg[ \tau'(z) R(z) \Bigg]_- 
         + \Bigg[ \widetilde{M}(z) A'(z) \Bigg]_-,
           \label{matchingT2}
    \eea
  where we have substituted $R(z) = - \widetilde{R}(z)$.
  At this stage, we can see that the field-dependent gauge kinetic term {\it does} affect 
  the quantum chiral ring relation 
    \footnote{The author thanks Ken Intriligator for a useful comment on this point.}
  :
  the second terms in the right hand sides denote that $T(z)$ and $\widetilde{T}(z)$ are affected 
  by the gauge kinetic terms.
  Indeed, in large $z$, the first equation of (\ref{matchingT2}) becomes
    \bea
    \left< W'(\Phi) \right> - \frac{i}{2} \left< \tau'(\Phi) \CW^\alpha \CW_\alpha \right>
    + \left< \bar{Q} Q \right> 
     =     0,
    \eea
  and this is the usual F-term equation.
  In the classical vacua, the second term does not contribute, but it does in the quantum vacua.
  
  Let us consider the effect of the second terms in (\ref{matchingT2}).
  In the constant $\tau$ case, by the duality map (\ref{matchingT}), 
  the first equation of (\ref{matchingT2}) should reduce to the second equation, as noted above.
  In the case with $\tau(\phi)$, 
  since the only difference between the second terms in (\ref{matchingT2}) is the sign, 
  they do not change the duality map (\ref{matchingT}).

%%%%%%%%%%%%%%%%%%%%%%%%%%%%%%%%%%%%%%%%%%%%%%%%%%%%%%%%%%%%%%%%%%%%%%%%%%%%%%%%%%%%%%%%%%%  acknowledgements
\section*{Acknowledgements}
  The author especially thanks Ken Intriligator for carefully reading the manuscript
  and giving useful comments.
  The author thanks Philip Argyres, Hiroshi Itoyama, Yutaka Ookouchi, Luca Mazzucato, Al Shapere, 
  Ta-Sheng Tai, Masato Taki, Seiji Terashima, Cumrun Vafa and Futoshi Yagi
  for useful discussions and helpful comments.
  The author also thanks Brown University, Harvard University, University of California, San Diego, 
  University of Cincinnati, University of Kentucky 
  and Perimeter Institute for Theoretical Physics for the hospitality during part of this project.
  The research of the author is supported in part by JSPS Research Fellowships for Young Scientists.

%%%%%%%%%%%%%%%%%%%%%%%%%%%%%%%%%%%%%%%%%%%%%%%%%%%%%%%%%%%%%%%%%%%%%%%%%%%%%%%%%%%%%%%%%%%%%%%%%%%%%
%\appendix

%\section*{Appendix}
%\section{}

%%%%%%%%%%%%%%%%%%%%%%%%%%%%%%%%%%%%%%%%%%%%%%%%%%%%%%%%%%%%%%%%%
%%% references
%%%%%%%%%%%%%%%%%%%%%%%%%%%%%%%%%%%%%%%%%%%%%%%%%%%%%%%%%%%%%%%%%


\begin{thebibliography}{99}

\bibitem{Vafa}
  C.~Vafa,
  %``Superstrings and topological strings at large N,''
  J.\ Math.\ Phys.\  {\bf 42} (2001) 2798
  [arXiv:hep-th/0008142].

\bibitem{CIV}
  F.~Cachazo, K.~A.~Intriligator and C.~Vafa,
  %``A large N duality via a geometric transition,''
  Nucl.\ Phys.\  B {\bf 603} (2001) 3
  [arXiv:hep-th/0103067].

\bibitem{CFIKV}
  F.~Cachazo, B.~Fiol, K.~A.~Intriligator, S.~Katz and C.~Vafa,
  %``A geometric unification of dualities,''
  Nucl.\ Phys.\  B {\bf 628} (2002) 3
  [arXiv:hep-th/0110028].
  
\bibitem{DV}
  R.~Dijkgraaf and C.~Vafa,
  %``Matrix models, topological strings, and supersymmetric gauge theories,''
  Nucl.\ Phys.\  B {\bf 644} (2002) 3
  [arXiv:hep-th/0206255];
%
%  R.~Dijkgraaf and C.~Vafa,
  %``On geometry and matrix models,''
  Nucl.\ Phys.\  B {\bf 644} (2002) 21
  [arXiv:hep-th/0207106];
%  
%  R.~Dijkgraaf and C.~Vafa,
  %``A perturbative window into non-perturbative physics,''
  [arXiv:hep-th/0208048].
  
\bibitem{DGLVZ}
  R.~Dijkgraaf, M.~T.~Grisaru, C.~S.~Lam, C.~Vafa and D.~Zanon,
  %``Perturbative computation of glueball superpotentials,''
  Phys.\ Lett.\  B {\bf 573} (2003) 138
  [arXiv:hep-th/0211017].
  
\bibitem{CDSW}
  F.~Cachazo, M.~R.~Douglas, N.~Seiberg and E.~Witten,
  %``Chiral rings and anomalies in supersymmetric gauge theory,''
  JHEP {\bf 0212} (2002) 071
  [arXiv:hep-th/0211170].

\bibitem{Seiberg2}
  N.~Seiberg,
  %``Adding fundamental matter to 'Chiral rings and anomalies in supersymmetric
  %gauge theory',''
  JHEP {\bf 0301} (2003) 061
  [arXiv:hep-th/0212225].
  
\bibitem{CSW2}
  F.~Cachazo, N.~Seiberg and E.~Witten,
  %``Chiral rings and phases of supersymmetric gauge theories,''
  JHEP {\bf 0304} (2003) 018
  [arXiv:hep-th/0303207].
  
\bibitem{IM}
  H.~Itoyama and K.~Maruyoshi,
  %``Deformation of Dijkgraaf-Vafa Relation via Spontaneously Broken N=2
  %Supersymmetry,''
  Phys.\ Lett.\  B {\bf 650} (2007) 298
  [arXiv:0704.1060 [hep-th]];
  
%\bibitem{Maruyoshi:2007hg}
  K.~Maruyoshi,
  %``Effective superpotential and partial breaking of N=2 supersymmetry,''
  arXiv:0710.2154 [hep-th];
  
%\bibitem{IM2}
  H.~Itoyama and K.~Maruyoshi,
  %``Deformation of Dijkgraaf-Vafa Relation via Spontaneously Broken N=2
  %Supersymmetry II,''
  Nucl.\ Phys.\  B {\bf 796} (2008) 246
  [arXiv:0710.4377 [hep-th]].
  
\bibitem{Ferrari}
  F.~Ferrari,
  %``Extended N=1 super Yang-Mills theory,''
  JHEP {\bf 0711} (2007) 001
  [arXiv:0709.0472 [hep-th]].
  
\bibitem{FIS}
  K.~Fujiwara, H.~Itoyama and M.~Sakaguchi,
  %``Supersymmetric U(N) gauge model and partial breaking of N = 2
  %supersymmetry,''
  Prog.\ Theor.\ Phys.\  {\bf 113} (2005) 429
  [arXiv:hep-th/0409060];
%
%\bibitem{FIS2}
%  K.~Fujiwara, H.~Itoyama and M.~Sakaguchi,
  %``Partial breaking of N = 2 supersymmetry and of gauge symmetry in the  U(N)
  %gauge model,''
  Nucl.\ Phys.\  B {\bf 723} (2005) 33
  [arXiv:hep-th/0503113];
%  
%\bibitem{FIS3}
%  K.~Fujiwara, H.~Itoyama and M.~Sakaguchi,
  %``Partial supersymmetry breaking and N = 2 U(N(c)) gauge model with
  %hypermultiplets in harmonic superspace,''
  Nucl.\ Phys.\  B {\bf 740} (2006) 58
  [arXiv:hep-th/0510255].

\bibitem{IMS}
  H.~Itoyama, K.~Maruyoshi and M.~Sakaguchi,
  %``N=2 Quiver Gauge Model and Partial Supersymmetry Breaking,''
  Nucl.\ Phys.\  B {\bf 794} (2008) 216
  [arXiv:0709.3166 [hep-th]].
  
\bibitem{ABSV}
  M.~Aganagic, C.~Beem, J.~Seo and C.~Vafa,
  %``Extended Supersymmetric Moduli Space and a SUSY/Non-SUSY Duality,''
  arXiv:0804.2489 [hep-th].
  
\bibitem{OT}
  K.~Ohta and T.~S.~Tai,
  %``Extended MQCD and SUSY/non-SUSY duality,''
  JHEP {\bf 0809} (2008) 033
  [arXiv:0806.2705 [hep-th]].
  
\bibitem{Maruyoshi}
  K.~Maruyoshi,
  %``SUSY/Non-SUSY Duality in U(N) Gauge Model with Partially Broken N=2
  %Supersymmetry,''
  Nucl.\ Phys.\  B {\bf 809} (2009) 279
  [arXiv:0808.2520 [hep-th]].
  
\bibitem{CKV}
  F.~Cachazo, S.~Katz and C.~Vafa,
  %``Geometric transitions and N = 1 quiver theories,''
  arXiv:hep-th/0108120.

\bibitem{OTatar}
  K.~h.~Oh and R.~Tatar,
  %``Duality and confinement in N = 1 supersymmetric theories from geometric
  %transitions,''
  Adv.\ Theor.\ Math.\ Phys.\  {\bf 6} (2003) 141
  [arXiv:hep-th/0112040].
  
\bibitem{KlebanovS}
  I.~R.~Klebanov and M.~J.~Strassler,
  %``Supergravity and a confining gauge theory: Duality cascades and
  %chiSB-resolution of naked singularities,''
  JHEP {\bf 0008} (2000) 052
  [arXiv:hep-th/0007191].
  
\bibitem{Kutasov}
  D.~Kutasov,
  %``A Comment on duality in N=1 supersymmetric nonAbelian gauge theories,''
  Phys.\ Lett.\  B {\bf 351} (1995) 230
  [arXiv:hep-th/9503086].

\bibitem{ASY}
  O.~Aharony, J.~Sonnenschein and S.~Yankielowicz,
  %``Flows and duality symmetries in N=1 supersymmetric gauge theories,''
  Nucl.\ Phys.\  B {\bf 449} (1995) 509
  [arXiv:hep-th/9504113].

\bibitem{KS}
  D.~Kutasov and A.~Schwimmer,
  %``On duality in supersymmetric Yang-Mills theory,''
  Phys.\ Lett.\  B {\bf 354} (1995) 315
  [arXiv:hep-th/9505004].

\bibitem{KSS}
  D.~Kutasov, A.~Schwimmer and N.~Seiberg,
  %``Chiral Rings, Singularity Theory and Electric-Magnetic Duality,''
  Nucl.\ Phys.\  B {\bf 459} (1996) 455
  [arXiv:hep-th/9510222].
  
\bibitem{EGKRS}
  S.~Elitzur, A.~Giveon, D.~Kutasov, E.~Rabinovici and A.~Schwimmer,
  %``Brane dynamics and N = 1 supersymmetric gauge theory,''
  Nucl.\ Phys.\  B {\bf 505} (1997) 202
  [arXiv:hep-th/9704104].
  
\bibitem{ABSV0}
  M.~Aganagic, C.~Beem, J.~Seo and C.~Vafa,
  %``Geometrically induced metastability and holography,''
  Nucl.\ Phys.\  B {\bf 789} (2008) 382
  [arXiv:hep-th/0610249].
  
\bibitem{ABF}
  M.~Aganagic, C.~Beem and B.~Freivogel,
  %``Geometric Metastability, Quivers and Holography,''
  Nucl.\ Phys.\  B {\bf 795} (2008) 291
  [arXiv:0708.0596 [hep-th]].

\bibitem{Seiberg}
  N.~Seiberg,
  %``Electric - magnetic duality in supersymmetric nonAbelian gauge theories,''
  Nucl.\ Phys.\  B {\bf 435} (1995) 129
  [arXiv:hep-th/9411149].

\bibitem{Ferrarif}
  F.~Ferrari and V.~Wens,
  %``Flavors in the microscopic approach to N=1 gauge theories,''
  arXiv:0904.0559 [hep-th].
  

  
  
\bibitem{DM}
  M.~R.~Douglas and G.~W.~Moore,
  %``D-branes, Quivers, and ALE Instantons,''
  arXiv:hep-th/9603167.
  
\bibitem{KN}
  I.~R.~Klebanov and N.~A.~Nekrasov,
  %``Gravity duals of fractional branes and logarithmic RG flow,''
  Nucl.\ Phys.\  B {\bf 574} (2000) 263
  [arXiv:hep-th/9911096];
%
%\bibitem{Polchinski:2000mx}
  J.~Polchinski,
  %``N = 2 gauge-gravity duals,''
  Int.\ J.\ Mod.\ Phys.\  A {\bf 16} (2001) 707
  [arXiv:hep-th/0011193].
  
\bibitem{LM}
  A.~Lawrence and J.~McGreevy,
  %``Local string models of soft supersymmetry breaking,''
  JHEP {\bf 0406} (2004) 007
  [arXiv:hep-th/0401034].

\bibitem{GMT}
  L.~Girardello, A.~Mariotti and G.~Tartaglino-Mazzucchelli,
  %``On supersymmetry breaking and the Dijkgraaf-Vafa conjecture,''
  JHEP {\bf 0603} (2006) 104
  [arXiv:hep-th/0601078].
  
\bibitem{EGK}
  S.~Elitzur, A.~Giveon and D.~Kutasov,
  %``Branes and N = 1 duality in string theory,''
  Phys.\ Lett.\  B {\bf 400} (1997) 269
  [arXiv:hep-th/9702014].

\bibitem{OV}
  H.~Ooguri and C.~Vafa,
  %``Geometry of N = 1 dualities in four dimensions,''
  Nucl.\ Phys.\  B {\bf 500} (1997) 62
  [arXiv:hep-th/9702180].
  
\bibitem{BRP}
  C.~E.~Beasley and M.~R.~Plesser,
  %``Toric duality is Seiberg duality,''
  JHEP {\bf 0112} (2001) 001
  [arXiv:hep-th/0109053].

\bibitem{FHHU}
  B.~Feng, A.~Hanany, Y.~H.~He and A.~M.~Uranga,
  %``Toric duality as Seiberg duality and brane diamonds,''
  JHEP {\bf 0112} (2001) 035
  [arXiv:hep-th/0109063].
  
\bibitem{IS}
  K.~A.~Intriligator and N.~Seiberg,
  %``Duality, monopoles, dyons, confinement and oblique confinement in
  %supersymmetric SO(N(c)) gauge theories,''
  Nucl.\ Phys.\  B {\bf 444} (1995) 125
  [arXiv:hep-th/9503179].
  
\bibitem{Mazzucato}
  L.~Mazzucato,
  %``Chiral rings, anomalies and electric-magnetic duality,''
  JHEP {\bf 0411} (2004) 020
  [arXiv:hep-th/0408240];
%
%\bibitem{Mazzucato:2005fe}
  L.~Mazzucato,
  %``Remarks on the analytic structure of supersymmetric effective actions,''
  JHEP {\bf 0512} (2005) 026
  [arXiv:hep-th/0508234].
  
\end{thebibliography}
\end{document}